\documentclass[10pt]{article}
\usepackage[english]{babel}

\usepackage[utf8]{inputenc}										
\usepackage[T1]{fontenc}										
\usepackage[margin=1in]{geometry}

\usepackage{amsmath,amsfonts,amssymb,amsthm,cancel,siunitx,
calculator,calc,mathtools,empheq,latexsym}

\usepackage{subfig,epsfig,tikz,float}		            % 

\usepackage{setspace}
\usepackage{booktabs,multicol,multirow,tabularx,array}          
\usepackage{natbib}
\title{\normalsize\bf%
\uppercase{Deep Weighted Monte Carlo: a hybrid option pricing framework using neural networks}
}

\author{%
Sándor Kunsági-Máté$^{*}$, Gábor Fáth, István Csabai, Gábor Molnár-Sáska
}
\begin{document}

\date{}

\maketitle

\vspace{-0.5cm}

\begin{center}
{\footnotesize 
ELTE Eötvös Loránd University, Budapest, Hungary\\
*Corresponding author\\
E-mail: skunsagimate@student.elte.hu\\
}
\end{center}

% -------------------------------------------------------------------
% Abstract
\bigskip
\noindent
{\small{\bf ABSTRACT.}
Recent studies have demonstrated the efficiency of Variational Autoencoders (VAE) to compress high-dimensional implied volatility surfaces into a low dimensional representation. Although this method can be effectively used for pricing vanilla options, it does not provide any explicit information about the dynamics of the underlying asset. In our work we present an effective way to overcome this problem. We use a Weighted Monte Carlo approach to first generate paths from a simple {\em a priori} Brownian dynamics, and then calculate path weights to price options correctly. We develop and successfully train a neural network that is able to assign these weights directly from the latent space. Combining the encoder network of the VAE and this new "weight assigner" module, we are able to build a dynamic pricing framework which cleanses the volatility surface from irrelevant noise fluctuations, and then can price not just vanillas, but also exotic options on this idealized vol surface. This pricing method can provide relative value signals for option traders.  

Key findings of the paper:
\begin{itemize}
    \item We confirmed the success of Variational Autoencoders in compressing volatility surfaces into low dimensional latent spaces in the swaption market.
    \item We successfully extracted the physically meaningful information from the latent representation of the market by translating the latent coordinates into weights of Monte Carlo paths.
    \item We developed a hybrid approach by combining Deep Learning and Weighted Monte Carlo that can price any type of option consistently with the latent model represented by the fully trained VAE. 
\end{itemize}
}


\baselineskip=\normalbaselineskip
% -------------------------------------------------------------------

\section{Introduction}\label{sec:intro}

In the early ‘70s, the Black-Scholes model (\citet{black1973}) put the pricing of options on a rigorous conceptual and mathematical basis. Assuming that the underlying asset volatility is constant was a reasonable simplification to make, and indeed, it was a good approximation for equity markets before the 1987 Black Monday crash. Since then the empirical observation is that options at different strike prices (K) imply different volatilities. This behaviour of the market is called the volatility skew or volatility smile. During the ‘90s several local volatility models appeared which were generalisations of Black-Scholes and tried to capture the observed volatility smile (see e.g.\ \citet{dupire}, \citet{derman}, \citet{dumas}). While these models could well reproduce the static pattern of the smiles they were unable to predict their dynamics. An effective solution to this problem was provided by stochastic volatility models (\citet{heston}, \citet{kim}, \citet{omar}). One frequently used approach proposed by Hagan and coworkers is the SABR stochastic volatility model (\cite{hagan2002}). SABR uses two coupled stochastic differential equations to describe the time development of the underlying asset price and its volatility. However, the dynamics of the volatility smile is still not described properly by the SABR model either. Traders have to recalibrate relatively often the SABR parameters (especially $\rho$ and $\nu$) to fit the model to observed data. 

As in the case of many fields of science, models used in the financial industry became more and more complicated, having larger and larger degrees of freedom to describe a specific problem with increasing accuracy. The usual model development process consists of two main steps: 1) Find the most relevant parameters of the problem and create a mathematical model that describes the relation between them; 2) Compare the model behaviour with the observations, and go back to step 1 if there is a non-negligible discrepancy. This clearly shows that during this iterative process it is really hard to find the number of minimally required parameters and the correct relations among them. 

Neural networks made a big impact on several areas of sciences and everyday life in recent years, and brought a completely different approach to solve problems. Deep Learning provides a data-driven approach where the model is adapted directly to the observed data. We do not have to care about all possible events anymore because the neural network can itself learn the complex relation between the real parameters, hence providing a better representation of the problem. Neural networks are flexible non-linear functions that can be trained to find an adequate low dimensional representation behind the data sometimes called the {\em active subspace} (\cite{deepuq}). 

There are two main types of applications of neural networks in option pricing: \textit{model replication} and \textit{model calibration}. Several stochastic models do not have an explicit formula for option prices, and therefore we need to run computationally expensive Monte Carlo simulations. Neural networks are able to find the connection between the model parameters and the outcome, and hence they can \textit{replicate} the model providing an approximation function that can be evaluated immediately (\citet{liu}, \citet{pagnottoni}, \citet{gaspar}). Another common problem is the calibration of the stochastic models to the observed market. In practice, finding the best fitting pricing model means to run again the time-consuming Monte Carlo simulation for several different sets of model parameters. Neural networks are also able to predict the model parameters directly from the observations (\citet{horvath}, \citet{bayer}), hence they reduce the duration of the calibration process by several orders of magnitude. However, this kind of model replication is only a speed-up, it is not able to alleviate systematic problems of the model if it is not particularly appropriate to the actual market it aims to describe. One may wonder if it might be more reasonable to \textit{apply the neural network model directly} to the observed data.

Recently, \citet{bergeron} demonstrated that Variational Autoencoders (VAEs) (\citet{kingma}) are able to find a good low dimensional representation of a large variety of volatility surfaces. They analyzed the FX market where they combined data from several currency pairs during the training process. They showed that a sufficiently trained VAE is not only able to interpolate between points on the volatility surface but in the case of sparse market data the decoder part of the VAE can efficiently reconstruct those missing points and provide a reliable extended surface. Although the results are very promising the trained neural network still remains a black box model which is relatively hard to interpret. Unlike the traditional approach, when we fit an analytic (e.g. SABR) model on the observed market and get the dynamics of the underlying asset, in case of VAEs we do not have any information about the stochastic processes under the hood. However, this information is crucial when we need to price more complex, path-dependent options (such as barriers). In our work we provide a hybrid solution for this problem where we combine Deep Learning with a method called Weighted Monte Carlo (\citet{avellaneda}).

This paper is organised as follows: In Section \ref{sec:data} we provide the details of our financial data and the conventional pricing methodology typically used by market participants. In Section \ref{sec:methods} we explain our methodology as well as the implemented model architecture. We also show the results of our approach made on the training and test sets. In Section \ref{sec:discussion} we discuss the key messages of our study, and finally in Section \ref{sec:conclusions} we summarize our work.

\section{Financial data and traditional pricing methodology}\label{sec:data}

A swaption (also known as a swap option) is an option to enter into an interest rate swap. Swaptions are used on the market to manage interest rate risk, either to hedge interest rate exposures or to trade on a directional view. A swaption has two key properties: the expiry and the tenor. Expiry defines the time interval in which one can enter into a swap, and tenor is the duration of the underlying swap. Expiries can range from a few months to 30+ years, while tenors are in the range of 1-30 years. This shows that there is a large variety of swaptions. To keep our analysis easy to interpret we chose a specific type of swaption with a 2-year tenor. We used historical data of GBP swaptions from ICAP (www.icap.com) with implied normal volatility values at 7 strike points around the at-the-money (ATM) point: ATM + [-50, -25, -12.5, 0.0, 12.5, 25, 50] basis points shifts, and at 7 expiries: 9M, 1Y, 18M, 2Y, 3Y, 4Y and 5Y. The observed time interval is between 2015 September 1 and 2020 October 21, hence we have volatility surface data for 1338 business days.

The traditional approach (Standard Model) to valuate swaptions uses a variant of Black's pricing formula originally developed to price options on futures \citet{black1976}. For swaptions Black's approach is justified if we work in the {\em annuity measure} (\citet{andersen}). This uses the annuity factor
\begin{equation}
    A_{k,m}(t) = \sum_{n=k}^{k+m-1} P(t, T_{n+1})(T_{n+1}-T_n)
\label{eq:annuity}
\end{equation}
as the numeraire. Here $t$ is the time of observation, $T_k$ is the starting date and $T_{k+m}$ is the maturity date of the (forward starting) swap. $P(t, T)$ are zero coupon bond prices (discount factors) defined for the underlying swap fixing dates. 

The requirement of no arbitrage connects the {\em forward swap rate} $S_{k,m}(t)$, i.e., the par rate of the forward starting swap, to the term structure $P(t,T)$ observable at the time of pricing (\citet{andersen}):
\begin{equation}
    S_{k,m}(t) = \frac{P(t, T_k)-P(t, T_{k+m})}{A_{k,m}(t)}.
\label{eq:par_rate}
\end{equation}
At time $t=0$, the forward swap rate $S_0=S_{k,m}(0)$ is known, but future interest rates are uncertain, so for any $t>0$ the swap rate $S_{k,m}(t)$ is a random process. Lack of arbitrage only dictates that in the annuity measure the forward swap rate should be a martingale (see \citet{andersen}):
\begin{equation}
    S_{k,m}(t_1) = E_{t_1}^A[S_{k,m}(t_2)], \quad \forall t1<t2.
\label{eq:swap_rate_martingale}
\end{equation}
Any model that makes the swap rate a martingale is arbitrage free and can be used to price other derivatives. In particular, in the annuity measure the {\em European swaption} price can be expressed as
\begin{equation}
    V_{k,m}(0) = N A_{k,m}(0)E_0^A[S_{k,m}(T_k)-K]
\label{eq:swaption_price}
\end{equation}
where $N$ is the notional of the swap, $T_k$ is the maturity of the swaption (the same as the starting time of the swap), $K$ is the strike rate, and expectation should be calculated under the annuity measure. If we assume that $S_{k,m}(T_k)$ is lognormally distributed we arrive at the annuity scaled Black formula typically used by practitioners. Note that lognormality can arise from the assumption that the swap rate follows a geometric Brownian motion, but to price a European option there is no need to know the dynamics in detail, only the marginal distribution counts. Similarly, if we assume a normal marginal (e.g., stemming from an arithmetic Brownian dynamics) we arrive at a Bachelier type pricing formula (\citet{choi2022}):
\begin{align}
    c_{payer} &= A(0)(S_0-K) \Phi(d) + \sigma \sqrt{T - t} \phi(d) & \label{eq:bachelier_call}\\ 
    c_{receiver} &= A(0)(K-S_0) \Phi(-d) + \sigma \sqrt{T - t} \phi(d), & \text{where} \quad d = \frac{S_0-K}{\sigma\sqrt{T - t}}. \label{eq:bachelier_put}
\end{align}
containing a normal implied volatility $\sigma$, and using $\Phi()$ and $\phi()$ as the probability distribution and density function of the standard normal distribution, respectively.  Note that payer/receiver are the swaption nomenclature for call/put options. Since some interest rates went negative on the market recently, the quoting convention for  swaptions typically works with normal implied volatilies. All vol surface data we obtained are quoted as normal implied volatilities, and should be interpreted in the Bachelier framework.

\iffalse
It is worth mentioning that for swaptions the ATM strike value, i.e., the $t=0$ forward swap rate $S_{k,m}(0)$ is different for each expiry $k$, since it is calculated according to the term structure. However, since we work in a "normal" (Bachelier) picture, option pricing is only sensitive to the strike to forward swap rate difference, and without loss of generality, we can set all ATM values to zero and search for possible paths that reproduce this market. Hence, in practice one should simply add the corresponding forward curve to the paths to calculate the price of other derivatives.
\fi

In the following we will use GBP swaptions with a 2-year tenor, but this is only an illustration; our method is expected to work with other markets as well. During our research we also investigated swaption products related to other expiries and tenors as well as EUR swaptions, where we got similar results to those presented here. Emphasis in this paper is on demonstrating the applicability of the proposed pricing method, and the GBP swaption market is used only as an example. 

\section{Methods and results}\label{sec:methods}

\subsection{Building a Variational Autoencoder}

After some investigation of the data we could clearly see that all volatility surfaces have a relatively similar shape which indicates that the 7x7=49 points are correlated and can be expressed with much less parameters. Similarly to the work of \citet{bergeron} we applied a Variational Autoencoder (VAE) to find the most compact latent representation behind the vol surface manifold. 

VAEs have been first introduced in \citet{kingma} and they are essentially the sophisticated version of Autoencoders. The input and output of these models are the same, hence they belong to the unsupervised model family meaning that they don't need labeled data for training. A simple Autoencoder neural network contains an encoder and a decoder model, where the encoder projects the input parameters (here the volatility surface) into a lower dimensional subspace, while the decoder tries to reproduce the original input parameters as accurately as possible given the latent representation. Although the reproduction efficiency of an Autoencoder can be high, there is no guarantee that the latent space is smooth enough meaning that two similar volatility surfaces are also close in the latent coordinates. This issue has been solved by the Variational Autoencoders in which the encoder projects the input on a distribution instead of a single point in the latent space and we force the distribution to be as close as possible to a multivariate normal distribution with a diagonal correlation matrix. In practice this discrepancy is measured by calculating the Kullback-Leibler divergence (also known as \textit{relative entropy}) between the two distributions. This extra loss term in the training process will help us to find the best latent representation of the data in which the latent coordinates are independent from each other (they do not correlate) and nearby latent data points represent similar volatility surfaces. 

The loss function we minimize is 
\begin{equation}
    \mathcal{L} = \mathcal{L}_{rec} + \beta \mathcal{L}_{KL},
\label{eq:vae_loss}
\end{equation}
where $\mathcal{L}_{rec}$ is the reconstruction loss between the real ($\mathbf{\sigma}$) and reconstructed ($\mathbf{\hat{\sigma}}$) volatility vectors and $\mathcal{L}_{KL}$ is the Kullback-Leibler divergence. We used the grid search approach to finetune the model hyperparameters over the $N_l - N_u$ plane, where $N_l$ and $N_u$ denote the number of hidden layers and the number of units in each layer, respectively, and searched for an optimal model in the $N_l \times N_u \in [2,3,4] \times [5,10,15,20]$ domain. With regard to the model complexity and accuracy trade-off, we finally selected the model whose key properties and training settings are summarized in Table \ref{table:vae_props}.
\newcolumntype{C}{>{\centering\arraybackslash}X}
\begin{table}[!htb]
\centering
\caption{Model architecture of the VAE as well as the training settings.}
\begin{tabularx}{1.0\textwidth}{ C | C }
\toprule
\midrule
Number of hidden layers (encoder) & 2 \\
Number of hidden layers (decoder) & 2 \\
Number of units in each layer & 10 \\
Activation function of hidden layers & elu \\
Batch size & 32 \\
Initial learning rate & 0.001 \\
Optimizer & Adam \\
$\beta$ & 5e-5 \\
\bottomrule
\end{tabularx}
\label{table:vae_props} 
\end{table}

\begin{figure}[h]
\includegraphics[width=1.\linewidth]{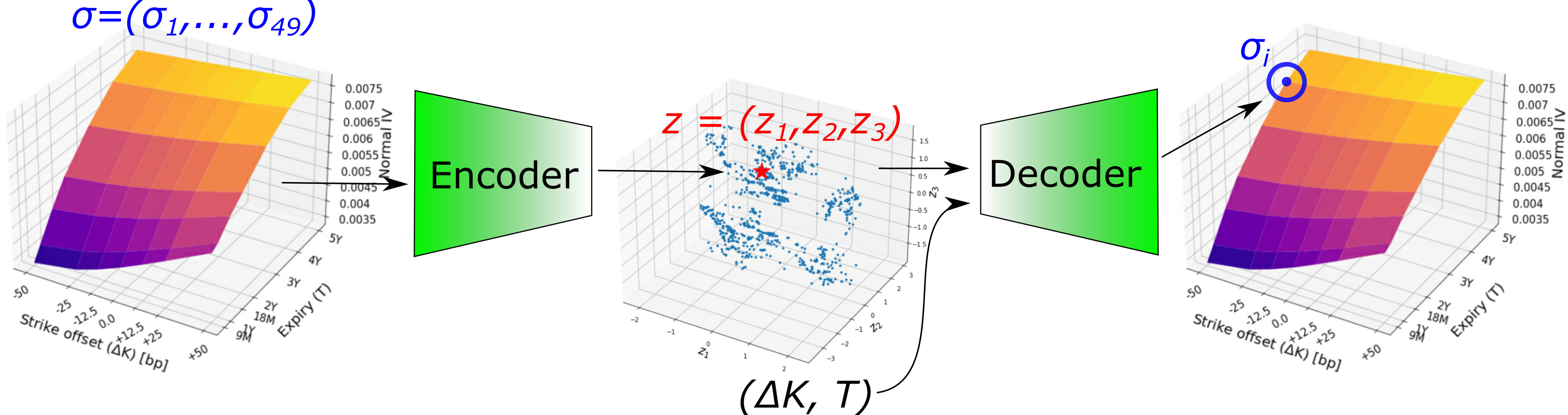}
\caption{Illustration of the data-flow through the Variational Autoencoder. From left to right: the encoder reads in and transforms the 49 points of the volatility surface into a 3 dimensional latent space; the decoder predicts a given point of the volatility surface using the latent representation of the volatility surface and the corresponding strike offset and expiry values.}
\label{fig:vae_illustration}
\end{figure}

As mentioned in \citet{bayer} there are two main approaches in the volatility surface reconstruction: the \textit{grid-based} and the \textit{pointwise} approaches. While in the grid-based approach the decoder predicts the 49 points of the volatility surface directly, in the pointwise approach the decoder is similar to a traditional pricer formula: it predicts a single implied volatility value from the model parameters (latent coordinates) and the $\Delta K$ strike offset and $T$ expiry. We chose the pointwise approach since it has two main advantages: 1) Once the VAE has been trained sufficiently, we can easily interpolate between the grid-points and can create arbitrary fine resolution of the volatility surface; 2) We can easily calculate the greeks using backpropagation, which is the most important parameter in hedging applications. The general scheme of the pointwise approach is shown in Figure \ref{fig:vae_illustration}.

\subsection{Training the Variational Autoencoder}

We split our data set into training and test sets using a split ratio of 7:3, where the two sets were separated chronologically at the date of 2019 April 5 to avoid leakage of future information in the training process. Hence, the blind test set contains normal as well as extreme (the pandemic crisis) market environments. According to our data analysis we defined the 2019-04-9 - 2020-01-15 and 2020-01-16 - 2020-10-21 time intervals as the \textit{normal} and \textit{extreme periods} of the test set, respectively. We also separated a validation set -- by randomly choosing 100 business days -- from the training set to continuously trigger the generalization performance during the training process and save only those model weights of the VAE that produce the lowest Mean Squared Error (MSE) in the validation set. This is a common practice to avoid model overfitting. Similarly to the work of \citet{bergeron} we tried out the latent dimensions of 2, 3 and 4. In Table \ref{table:mse_for_diff_lat_dims} we summarized the model reconstruction performance (expressed in the Mean Squared Error) on the training set with respect to each dimensionality of the latent space.

%\newcolumntype{C}{>{\centering\arraybackslash}X}
\begin{table}[!htb]
\centering
\caption{Mean Squared Error between the reconstructed and real volatility surfaces of the training set using different number of latent dimensions.}
\begin{tabularx}{1.0\textwidth}{ C | C | C | C }
\toprule
\midrule
Number of latent dimensions & 2 & 3 & 4\\
Mean Squared Error & 1.068e-4 & 6.152e-5 & 4.339e-5\\
\bottomrule
\end{tabularx}
\label{table:mse_for_diff_lat_dims} 
\end{table}

\begin{figure}[H]
\includegraphics[width=1.\linewidth]{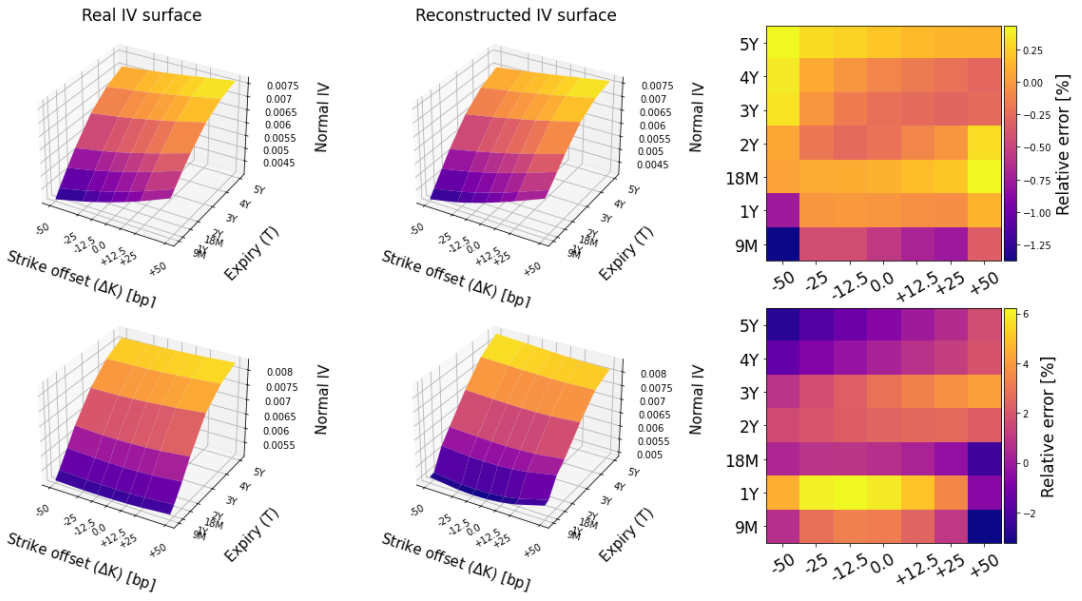}
\caption{Illustration of the implied volatility surface reproduction efficiency applying the fine tuned Variational Autoencoder network on the training set. The top and bottom rows correspond to the best and worst performing cases, respectively. From left to right we plotted the observed and reconstructed implied volatility surfaces as well as the relative error between them, respectively.}
\label{fig:rep_efficiency_illustration}
\end{figure}

As we can observe increasing the dimensions from two to three reduced the MSE by half. However, using 4 dimensions did not help significantly, therefore considering the trade-off between model complexity and performance we used three latent dimensions for our study. In Figure \ref{fig:rep_efficiency_illustration} we plotted the reconstruction efficiency of the VAE regarding to the training set in the best and worst cases considering the Mean Relative Absolute Error (MRAE) between the real ($\boldsymbol{\sigma}_{real}$) and reconstructed ($\boldsymbol{\sigma}_{rec}$) implied volatility surfaces (see Eq. \ref{eq:MRAE}).

\begin{equation}
    MRAE = \frac{1}{N} \sum\limits_{i=1}^N \frac{ \left| \sigma_{real,i} - \sigma_{rec,i} \right|}{\sigma_{real,i}} 
    \label{eq:MRAE}
\end{equation}

\begin{figure}[h]
\includegraphics[width=1.\linewidth]{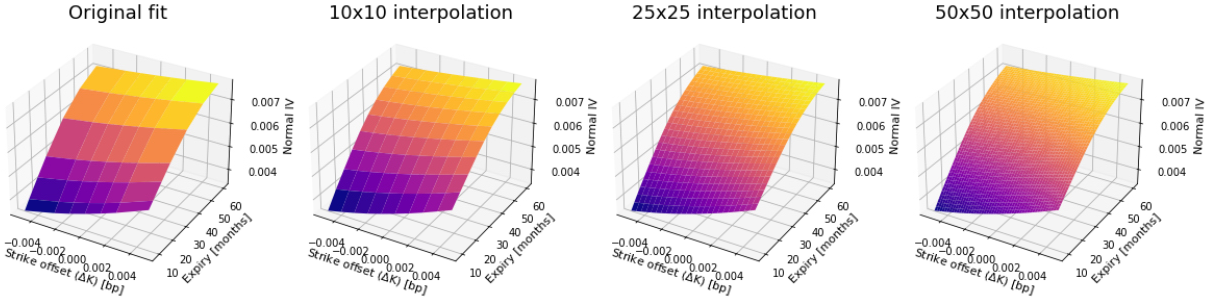}
\caption{Illustration of the interpolation capability of the decoder network. From left to right we increased the resolution from the original unequally spaced 7x7 grid to equally spaced 10x10, 25x25 and 50x50 grids.}
\label{fig:interp_illustration}
\end{figure}

We can observe that in the best case the relative error is below 1\% at almost every point of the strike offset - expiry grid, and even in the worst case the errors increase mostly to a few percents. This demonstrates that we found a model that can well describe the large variety of volatility surfaces in the training set with only three parameters. Furthermore, this model is able to reliably interpolate between the observed strike offset - expiry grid points. In Figure \ref{fig:interp_illustration} we demonstrate this capability of the decoder network by continuously increasing the resolution.

\begin{figure}[H]
\includegraphics[width=1.\linewidth]{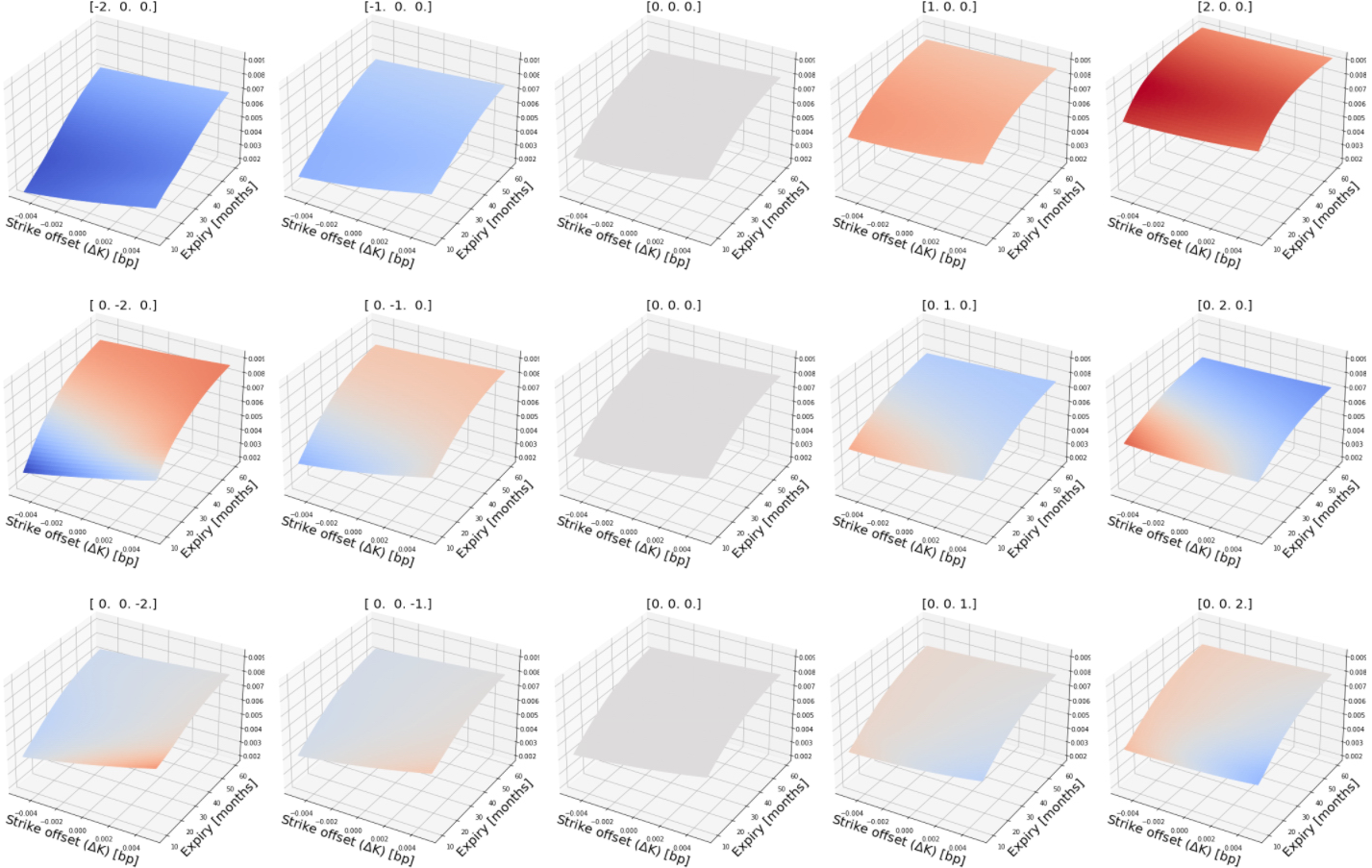}
\caption{Effect of each latent dimension on the volatility surface. Each row represents a movement in the latent space from -2 to +2 along the corresponding latent coordinate. The related position in the latent space can be seen in the title of each plot. Color coding represents the difference between the actual volatility surface and the volatility surface at the origin [0,0,0] of the latent space.}
\label{fig:eff_of_lat_dims}
\end{figure}

We can notice that the interpolated volatility surfaces remained smooth which means that we did not overfit the decoder network on the data. Now, let us investigate the effect of each latent dimension on the volatility surfaces, where we used the finest, 50x50 resolution (see Figure \ref{fig:eff_of_lat_dims}). 

We can see that each latent dimension has a clear meaning. Moving along the first latent dimension increases the height of the volatility surface. Modifying the second latent coordinate adjusts the skewness between the bottom left and top right corner of the volatility surfaces. Finally, the last latent coordinate has only a slight effect: it changes the slope of the volatility smiles related to the shortest expiries. 

Now, let us investigate the reproduction efficiency of the VAE on the blind test set. Before diving into the details, we would like to highlight again that the main purpose of training a Variational Autoencoder on historical data is to catch the correlations between the volatility surface points and find a low dimensional latent space from which the decoder neural network can reproduce the observed volatility surfaces the best. After the VAE is trained successfully, we can say that the network is good at \textit{interpolating} between the volatility surfaces of the training set. When we analyse the performance of the VAE on unseen data we are essentially interested in its \textit{extrapolation} capability that depends mainly on how reliable the latent space is. 

There are two ways to get the latent representation of a given volatility surface: 1) We can use the encoder network to map directly to the latent parameters (we will call this "direct encoding") or 2) we can fit the decoder network using an optimization algorithm searching in the latent space (to be called "calibration"). Direct encoding is certainly faster, but we have seen cases when the encoder network failed to predict the correct latent coordinates. In many cases we could achieve a smaller reproduction error using the calibration approach. The observed effect may originate from the fact that we intentionally introduce some noise in the training process and therefore omitting the encoder will reduce the reconstruction error. However, deeper analysis of this phenomenon is out of the scope of the current work.

\begin{figure}[H]
\centering
\includegraphics[width=0.7\linewidth]{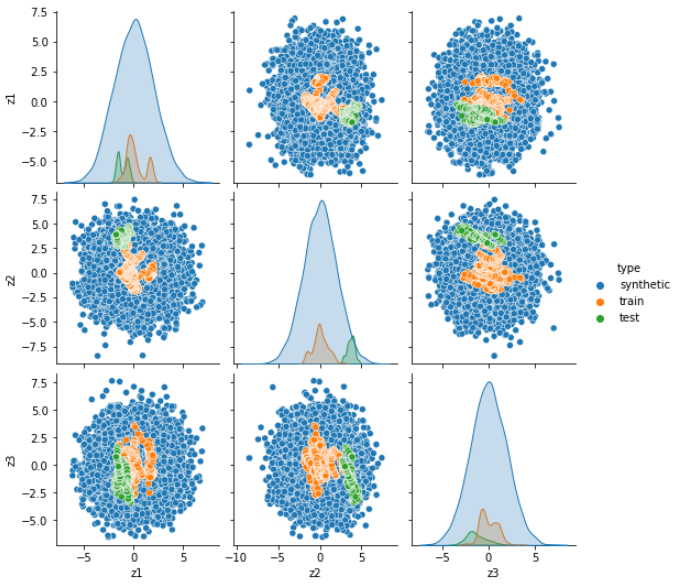}
\caption{Distribution of the synthetic, training and test set in the latent space.}
\label{fig:latent_space}
\end{figure}

To improve direct encoding we decided to boost training with synthetic data. Given that our latent space is designed to be normally distributed, we randomly sampled 10,000 points of the latent space from a Gaussian distribution with zero mean and a variance of 4 to cover a broader space than that of the training set (see Figure \ref{fig:latent_space}). We decoded these points into synthetic volatility surfaces,\footnote{We found 44 unrealistic volatility surfaces of the 10,000 that contained negative values. These data were removed from the training set.} and used this new synthetic data set to retrain the encoder part of the VAE. During this \textit{fine tuning} of the encoder, the decoder layers were kept fixed.

\begin{figure}[H]
\centering
\includegraphics[width=1.\linewidth]{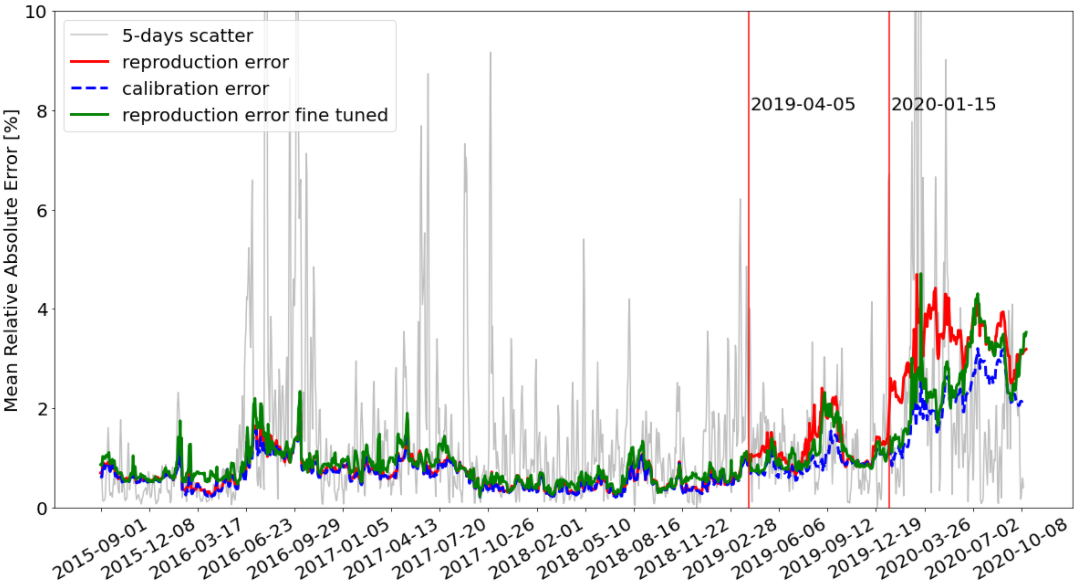}
\caption{Time evolution of the mean relative absolute error on the training and test sets. For reference we also show the standard deviation of volatility surfaces within a 5-day long sliding window (gray thin line).}
\label{fig:error_on_dataset}
\end{figure}

We determined the mean relative absolute error (MRAE) of the whole historical data set for both the original and fine tuned versions of direct encoding and also for the calibration approach. In this latter case we used the Trust Region Reflective algorithm of \citet{coleman}. In Figure \ref{fig:error_on_dataset} we plotted the time evolution of MRAE related to the three approaches, and we determined for reference the standard deviation of volatility surfaces from their mean value within a 5-days long sliding window (gray thin line). This reference quantity illustrates how large is the fluctuation in the market from day to day. We can observe that all approaches produce almost the same accuracy, 0.66-0.79\% in the training set. However, in the blind test set the MRAE curves start to deviate and the lowest values are produced by the optimization method (blue dashed line). Fine tuned direct mapping (green line) performs better at several time regions (e.g.: 2020-01-15 - 2020-06-30) than its simply trained version (red line), which means that the retraining process successfully increased the accuracy. In the normal period of the data set the MRAE is about 0.86--1.23\% while in the extreme period it increases to 2.14--3.30\%, which is still acceptable. We found that the trained model worked well for almost the whole first year of the unseen test set producing the same MRAE as in the training set, and only started to deviate at the beginning of the coronavirus crisis in the middle of March 2020. This good performance is an indication that the model is not overfitting on the training set, and therefore it is able to extrapolate for a relatively long time horizon with only three latent parameters. The increase of the error after one year is due to a drastic change in the market. The low dimensional submanifold which represents the physical market starts to move away from the \textit{active subspace} of the model learnt from the training set.

\subsection{Weighted Monte Carlo}

After successfully training the Variational Autoencoder on our swaption data we get a generative model with three model parameters that can describe with good accuracy the shape of every volatility surface that may typically arise on the market. Now imagine that we currently observe the volatility surface of swaptions today. Let's investigate how we can use the trained VAE on this new observation:

\begin{itemize}
    \item If we have all the points of the volatility surface, we can get the model parameters by simply transmitting the data through the encoder.
    \item If we have only a few points of the volatility surface we can still get the related latent representation of the market by calibrating the decoder on the data applying a given optimization method.
    \item Now using the decoder we can price any vanilla option related to a given ($\Delta K$, $T$) pair.
    \item We can also generate arbitrary number of synthetic, but realistic volatility surfaces.
\end{itemize}

The shortcoming of our description is that the decoder network predicts the implied volatility of the market {\em directly}, and does not give us any information about the dynamics of the underlying asset, which we would need to price any other non-vanilla derivatives on this market. Any information about the dynamics minimally required to describe the volatility surface is there implicitly in the latent coordinates. Therefore, to go beyond vanillas, we need to build another "decoder" neural network that is able to translate the latent representation of the volatility surface into the possible {\em dynamics} of the underlying asset that the market participants expect. For this purpose we used the so-called Weighted Monte Carlo method that was first introduced in the field of option pricing by \citet{avellaneda}. In this framework we start with an initial set of paths. We generated $\nu = 20,000$ paths having a 5-year time horizon with 25,000 timesteps. In case of regular Monte Carlo we take all paths with the same weight, $q_i = 1/\nu, i \in [1,...,\nu]$, into account. Given a security (in our case a swaption) with a path dependent payoff $g_i$, we can calculate its discounted price $\Pi_g$ as
\begin{equation}
    \Pi_g = \frac{1}{\nu} \sum\limits_{i=1}^{\nu} g_i.
    \label{eq:regmc}
\end{equation}

In the Weighted Monte Carlo approach we assign a unique weight to each path and modify them in a way that we can reproduce the observed market,
\begin{equation}
    \Pi_g = \sum\limits_{i=1}^{\nu} g_i p_i.
    \label{eq:wmc}
\end{equation}

This means however that we have a vast number of free parameters and therefore we should introduce some regularization. Similar to \citet{avellaneda} we used the relative entropy measure, $D(\mathbf{p}/\mathbf{q})$, between the $\mathbf{q}$ prior (here: the uniform distribution) and $\mathbf{p}$ posterior distributions
\begin{equation}
    D(\mathbf{p}/\mathbf{q}) = \sum\limits_{i=1}^{\nu} p_i \text{ln} \left( \frac{p_i}{q_i} \right).
    \label{eq:rel_ent}
\end{equation}

If we have observation of prices $c_j$ of $N$ securities, we can define a $\mathbf{G}$ payoff matrix with $\nu$ rows and $N$ columns
\begin{equation*}
\mathbf{G} = 
\begin{pmatrix}
g_{1,1} & g_{1,2} & \cdots & g_{1,N} \\
g_{2,1} & g_{2,2} & \cdots & g_{2,N} \\
\vdots  & \vdots  & \ddots & \vdots  \\
g_{\nu,1} & g_{\nu,2} & \cdots & g_{\nu,N}. 
\end{pmatrix}
\label{eq:payoff_mtx}
\end{equation*}
The calibration algorithm then goes for finding those $p_i$ values that satisfy
\begin{equation}
    \sum\limits_{i=1}^{\nu} p_i g_{i,j} = c_j, \quad j \in (1,2,...,N).
    \label{eq:prices}
\end{equation}
We can also rewrite Eq.\ \ref{eq:prices} in matrix form,
\begin{equation}
    \mathbf{p} \mathbf{G} = \mathbf{c}
    \label{eq:prices_mtx}
\end{equation}

In the traditional Weighted Monte Carlo approach we solve this problem using Lagrange multipliers $\lambda_j$, $j \in (1,2,...,N)$), and minimize the loss
\begin{equation}
    \mathcal{L} = D(\mathbf{p}/\mathbf{q}) + \sum\limits_{j=1}^N \lambda_j \left( \sum\limits_{i=1}^{\nu} p_i g_{i,j} - c_j \right).
    \label{eq:wmc_loss}
\end{equation}
According to \citet{avellaneda} the weights can be expressed with the Lagrange multipliers as
\begin{equation}
    p_i = \frac{1}{Z( \mathbf{\lambda} )} \exp \left( \sum\limits_{j=1}^{N} g_{i,j} \lambda_j \right), \quad \text{where}  \quad Z(\mathbf{\lambda}) = \frac{1}{\nu} \sum\limits_{i=1}^{\nu} \exp \left( \sum\limits_{j=1}^{N} g_{i,j} \lambda_j \right).
    \label{eq:probs}
\end{equation}

In this way the weight vector is a function $\mathbf{p} = f(\mathbf{\lambda})$ of as many parameters as observations we have on the market. In our case, we have 7x7=49 data points on the vol surface, which means that we can calculate the price of at least 49 securities. However, using the Variational Autoencoder we found that this 49-dimensional space is degenerate and the market can be described with good accuracy by only three parameters. Therefore we can try to build a new \textit{Weight Decoder} neural network, $\Phi_{WD}$, which is able to translate the latent representation of the market into the weights of possible paths,
\begin{equation}
    \mathbf{p} = \Phi_{WD}(\mathbf{z}).
    \label{eq:probdec}
\end{equation}

\begin{figure}[h]
\includegraphics[width=1.\linewidth]{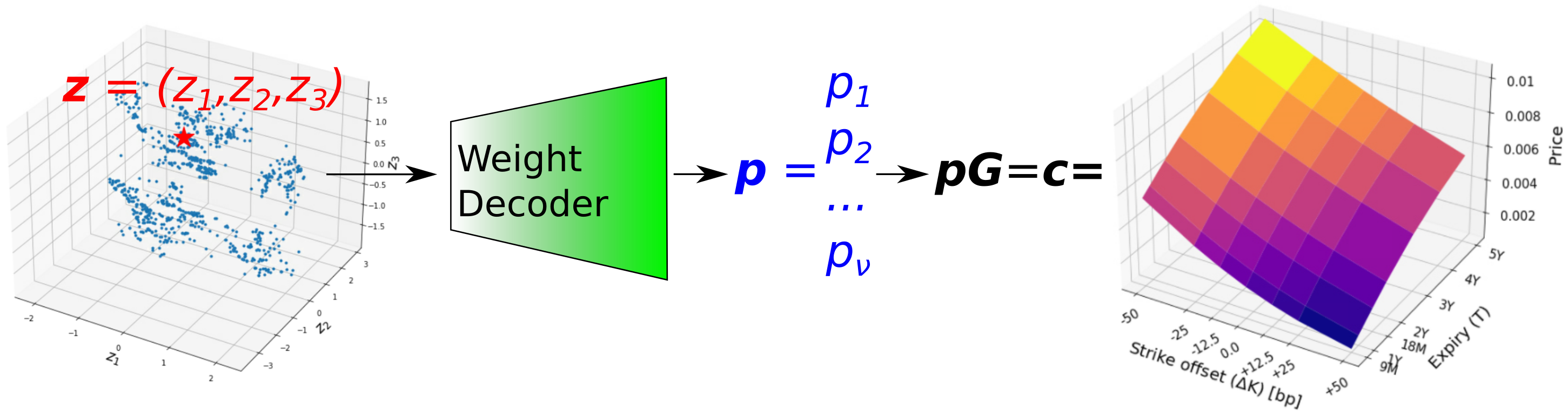}
\caption{Illustration of the data-flow through the Weight Decoder neural network. From left to right: the weight decoder reads in the latent coordinates and transforms them into the weights of different paths; we calculate then the prices of securities using Weighted Monte Carlo. During the training phase we try to reproduce the prices of securities (that we previously determined from the observed implied volatility values) as accurately as possible.}
\label{fig:probdec_illustration}
\end{figure}

In Figure \ref{fig:probdec_illustration} we illustrate the data processing from the latent space to the vanilla price surface using the Weight Decoder. Note that we work with the option price surface, and not the vol surface here, as the Monte Carlo method gives prices, not IVs directly.  

First we apply the Weight Decoder on the latent space to get the weights related to the paths. Then we calculate the prices of securities under consideration using Weighted Monte Carlo. Since we do not have an explicit formula from prices of securities to implied volatilities, we generated call ($c_{call}$) and put ($c_{put}$) options related to the observed volatility surface using the Bachelier formulae, Eqs.\  \ref{eq:bachelier_call} and \ref{eq:bachelier_put}. We ran the training process to reproduce these prices as accurately as possible.

It is worth mentioning that we use the call $\hat{c}_{call}$ and put prices $\hat{c}_{put}$ related to the reconstructed implied volatilities of the VAE decoder network. This is a convenient choice since this way we train the weight decoder on smooth price surfaces that do not contain noise anymore. The key properties of the model architecture is summarized in Table \ref{table:prob_dec_props}.

%\newcolumntype{C}{>{\centering\arraybackslash}X}
\begin{table}[!htb]
\centering
\caption{Model architecture of the weight decoder as well as the training settings.}
\begin{tabularx}{1.0\textwidth}{ C | C }
\toprule
\midrule
Number of hidden layers & 2 \\
Number of units in each layer & 20/49 \\
Activation function of hidden layers & elu \\
Activation function of the readout layer & softmax \\
Batch size & 32 \\
Initial learning rate & 0.01 \\
Optimizer & Adam \\
$\gamma$ & 1e-8 \\
\bottomrule
\end{tabularx}
\label{table:prob_dec_props} 
\end{table}

\subsection{Put-call parity}\label{seq:putcall}

To avoid arbitrage opportunities it is necessary to ensure that the put-call parity holds for all predicted prices:
\begin{equation}
    K - S_0+ c_{call} - c_{put} = \Delta K + c_{call} - c_{put} = 0 
    \label{eq:put_call_parity}
\end{equation}

Instead of introducing an extra loss term for the put-call parity we embedded this relation as a constraint into the put price estimation from call prices:

\begin{equation}
    \hat{\hat{c}}_{put} = \Delta K + \hat{\hat{c}}_{call} 
    \label{eq:put_price}
\end{equation}

The loss of call and put price reconstruction is then considered during the training, where we denoted the reconstructed call and put prices by $\hat{\hat{c}}_{call}$ and $\hat{\hat{c}}_{put}$, respectively. The final form of the loss function is
\begin{equation}
    \mathcal{L} = \mathcal{L}_{rec,call} + \mathcal{L}_{rec,put} + \gamma D(\mathbf{p}/\mathbf{q}),
\label{eq:prob_dec_loss}
\end{equation}
where $\mathcal{L}_{rec,call}$ and $\mathcal{L}_{rec,put}$ are the reconstruction losses of the prices of put and call options, $D(\mathbf{p}/\mathbf{q})$ is the relative entropy between the predicted weights and the uniform distribution, and $\gamma$ is a weighting constant.

\subsection{Martingale condition}

%\iffalse
The main purpose of the calibration method described above is to reproduce the observed option prices. However, to price derivatives without arbitrage, the paths have to satisfy the martingale condition Eq.\ \ref{eq:swap_rate_martingale}. Although the original paths have been generated from a martingale process, the re-weighting process can break the martingale condition as was found in \citet{elices}. They argue that in case of calibrating on several expiries with steep smiles, the resulting underlying process may fail to satisfy the martingale condition. 

\begin{figure}[H]
\centering
\includegraphics[width=1.\linewidth]{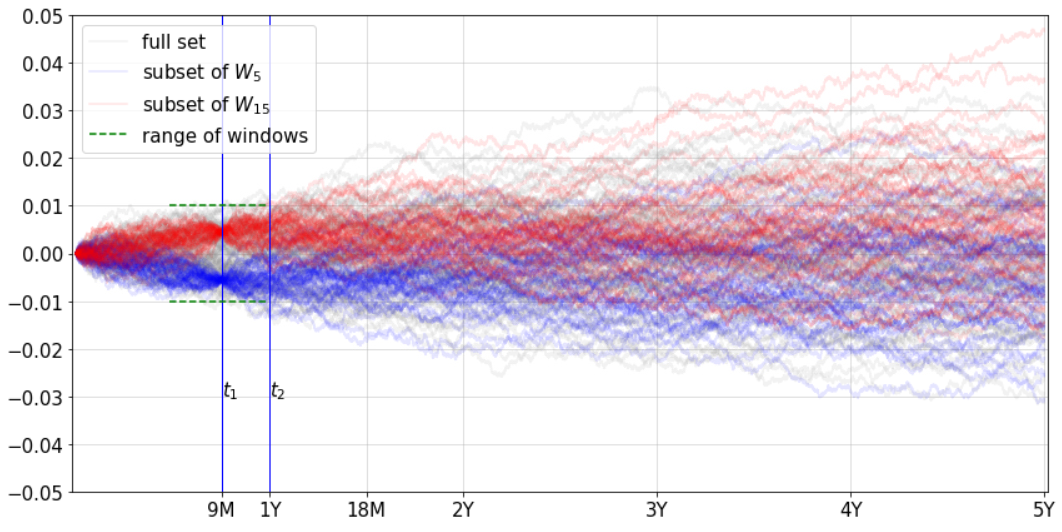}
\caption{Illustration of the martingale condition in the Monte Carlo environment. The blue and red paths correspond to two different windows, $W_{0.75}(-0.05)$ and $W_{0.75}(0.05)$ at time $t_1=0.75$. The grey paths represent the full path set and the green vertical lines denote the range within we defined the windows.}
\label{fig:mart_illustration}
\end{figure}

Note that including the put-call parity breaking penalty per Section \ref{seq:putcall} is already a way to curb deviations from the martingale condition. However to force this requirement more rigorously in a discrete Monte Carlo environment we would need more control. Consider the subset of paths that go through a narrow window at time $t$ around the point $S=s$, denoted $W_t(s) = [s-\Delta, s+\Delta]$, where $\Delta=0.001$ by our choice. In Figure \ref{fig:mart_illustration} we illustrate these paths going through two different windows. 

To be a martingale process, for every window $W=W_{t_1}(s)$ and later time $t_2$ the relationship
\begin{equation}
    E_{t_1}^A[ S_{t_1} | S_{t_1} \in W ] = E_{t_1}^A[ S_{t_2} | S_{t_1} \in W] 
    \label{eq:mart_window}
\end{equation}
should hold. This can be rewritten in a path discrete form
\begin{equation}
\sum\limits_{j|S_{j,t_1} \in W} \left( S_{j,t_2} - S_{j,t_1} \right) \frac{p_j}{\sum\limits_{i|S_{i,t_1} \in W} p_i} = 0.
\label{eq:mart_window_discrete}
\end{equation}

One way to introduce the martingale condition into the training process is to create additional fictitious derivatives with zero price ($c_{f} = 0$) that have the payoff function
\begin{equation}
 g_{\text{mart}, j} = \left( S_{j,t_2} - S_{j,t_1} \right), \quad \text{where} \quad S_{j,t_1} \in W
\label{eq:payoff_func_mart}
\end{equation}
for a set of windows $W_{t_i}(s)$ scattered around time and rate space. Considering only the consecutive expiries (as in \citet{elices}) using windows at 20 different positions between -0.01 and 0.01\footnote{Note that the range of strikes is between -0.005 and 0.005, hence we cover the important region.} we have 6x20=120 additional derivatives. During the training process we force the model to reproduce the zero prices of these derivatives by adding the same loss term as for the call and put options. 

\begin{figure}[H]
\centering
\includegraphics[width=1.\linewidth]{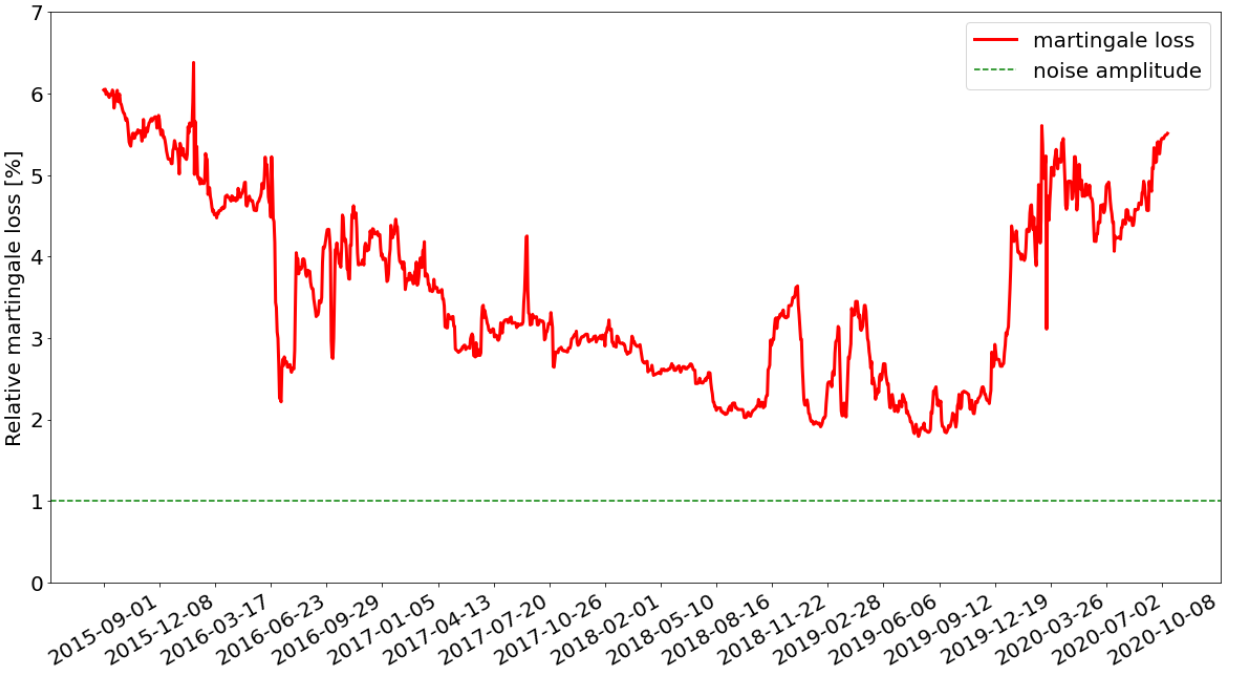}
\caption{Time evolution of the observed martingale loss (red line) relative to the range of window positions ([-0.01,0.01]). For reference we plotted the level of martingale loss (green dashed line) related to the noise originated from the path generation process.}
\label{fig:mart_loss}
\end{figure}

During our analysis we found that this extra loss term is \textit{computationally very expensive}, increasing the time of a single training step by an order and the training loss was still fluctuating, the model did not converge. Therefore, we decided to train the weight decoder without these loss terms and investigate only the deviation of our emerging measure from the martingale property. In Figure \ref{fig:mart_loss} we plotted the time evolution of the mean martingale loss relative to the support of window positions (0.01-(-0.01) = 0.02). In this way we give an intuition about the severity of the observed deviation from the martingale condition.

We can see that the discrepancy is relatively small and stays in the 2-6\% range for the whole data set. Note, however, that part of this noise comes from the Monte Carlo method itself. To estimate this contribution we regenerated 50 times the used set of 20,000 paths and calculated the standard deviation of the martingale loss for the unweighted paths. The relative value of this amplitude was found to be about 1\% (see green dashed line in Figure \ref{fig:mart_loss}). This means that a non-negligible part of the martingale loss comes from the sampling noise of the Monte Carlo method. We can conclude that using the put-call parity relation during the training process is sufficient to keep the martingale loss at a sufficient level. The addition of a more complex set of constraints makes the approach difficult to manage and may not have tangible benefits. One should however keep the option of these extra controls in mind if dealing with securities particularly sensitive to the martingale condition.

\subsection{Training the weight decoder}

\begin{figure}[H]
\centering
\includegraphics[width=1.\linewidth]{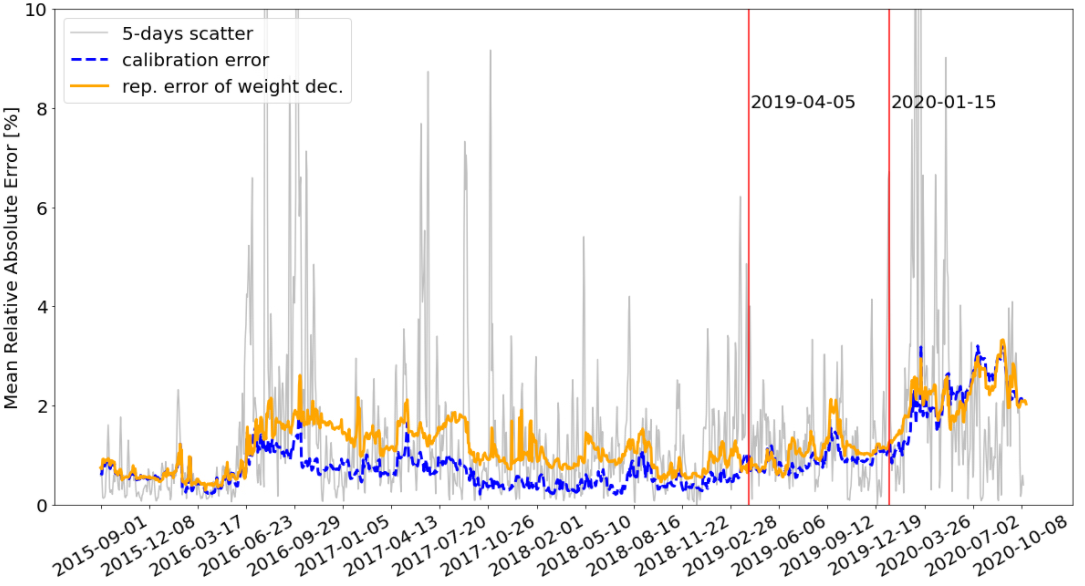}
\caption{Time evolution of the Mean Relative Absolute Error of the weight decoder related to the training and test sets (orange line). We compared our results to the best performing calibration method (blue dashed line). For reference we calculated the relative standard deviation of volatility surfaces within a 5-days long sliding window (gray thin line).}
\label{fig:error_on_dataset_weight_dec}
\end{figure}

At this stage we have a latent representation of the swaption market that contains all necessary information. Since the observable market essentially shows us the expectations of the market participants, the derived latent space must be in a direct relation with the possible paths of the underlying asset. Therefore we built a new neural network, the  \textit{weight decoder} that is able to translate the latent coordinates into the weights of the paths to be used later in the Weighted Monte Carlo method. We trained this model on the same training set as was used in the fine tuning process of the encoder network. To continuously trigger the performance of the network we chose the original training set (the period of 2015-09-01 - 2019-04-09) as the validation set. In Figure \ref{fig:error_on_dataset_weight_dec} we plot the reconstruction efficiency of the weight decoder, where we made a comparison with the best performing calibration method (see Fig. \ref{fig:error_on_dataset}). The reconstructed implied volatilities have been iteratively calculated from the calculated call price surface using the Bachelier formula.

We can observe that the reproduction efficiency achieved by the weight decoder is very close to the calibration method (1.06\% and 1.63\% for training and test sets, respectively). This means that the weight decoder predicts the correct weights from the latent space, that can be assigned to the paths to reproduce the observed market accurately. 

\begin{figure}[H]
\includegraphics[width=1.\linewidth]{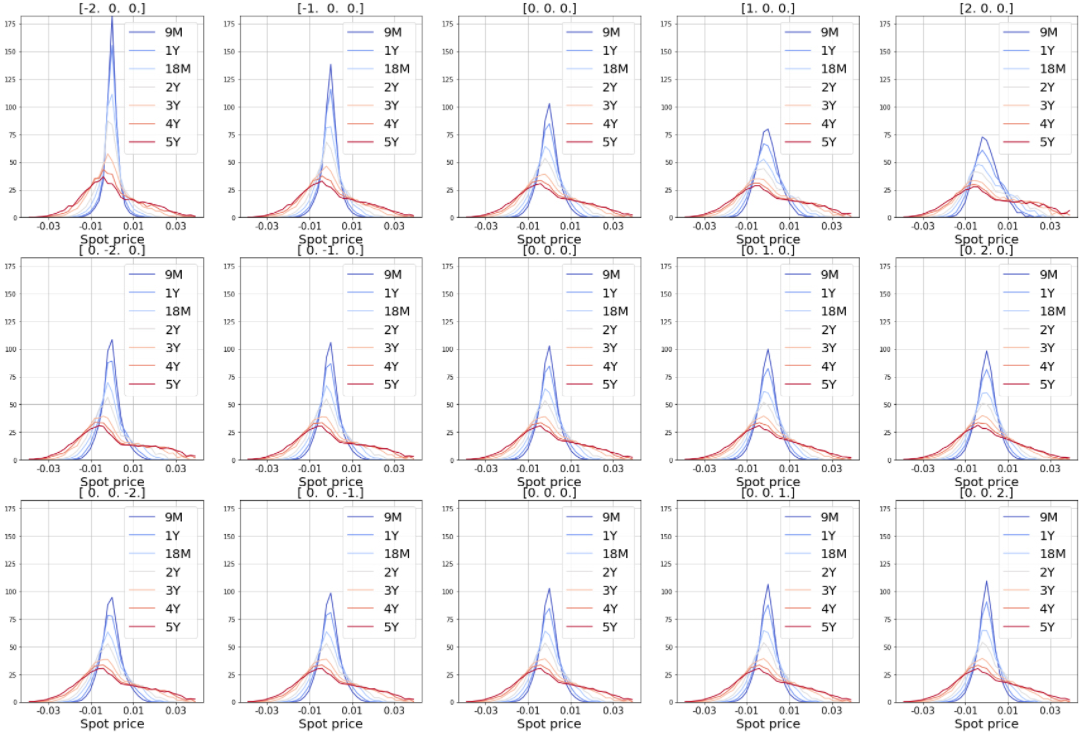}
\caption{Effect of each latent dimension on the risk neutral densities. Each row represents a movement in the latent space from -2 to +2 along the corresponding latent coordinate. The related position in the latent space can be seen in the title of each plot. All of the risk-neutral densities are scaled the same for comparison purposes.}
\label{fig:eff_of_lat_dims_on_rnd}
\end{figure}

In Figure \ref{fig:eff_of_lat_dims_on_rnd} we plotted the effect of each latent dimension on the risk-neutral densities using the trained weight decoder where we analyzed the same positions in the latent space as in Figure \ref{fig:eff_of_lat_dims}. We can observe that the first latent dimension has the most significant effect, that modifies the volatility of the paths. The other two coordinates have much less influence on the risk neutral densities as we expected from Figure \ref{fig:eff_of_lat_dims}. 

\subsection{Pricing a barrier call option}

In the following we will price a path-dependent option, namely an up barrier call option with an expiry of five years. The payoff value of this product is zero in case of paths that exceed the predefined barrier before the expiry and is equal to the price of a regular call option otherwise. We chose the swaption market of 2016-03-30, where we got an accurate fit using the Weight Decoder. For comparison we fitted the analytic $\beta=0$ SABR formula as well on the volatility smile related to the 5Y expiry\footnote{The calibration of stochastic models on the whole volatility surface is often impossible and therefore the usual way for calibrating e.g. the SABR model is to fit the approximation formula to each volatility smile, individually.} to get the $\alpha$, $\rho$ and $\nu$ model parameters driving the two stochastic differential equations 
\begin{align}
    dS_t &= \sigma_t S_t^{\beta} dW_t^1\label{eq:sabr1}\\ 
    d\sigma_t &= \nu \sigma_t dW_t^2, & \text{where} \quad dW_t^1 dW_t^2 = \rho dt \quad \text{and} \quad \alpha = \sigma_0 \label{eq:sabr2}
\end{align}

\begin{figure}[H]
\includegraphics[width=1.\linewidth]{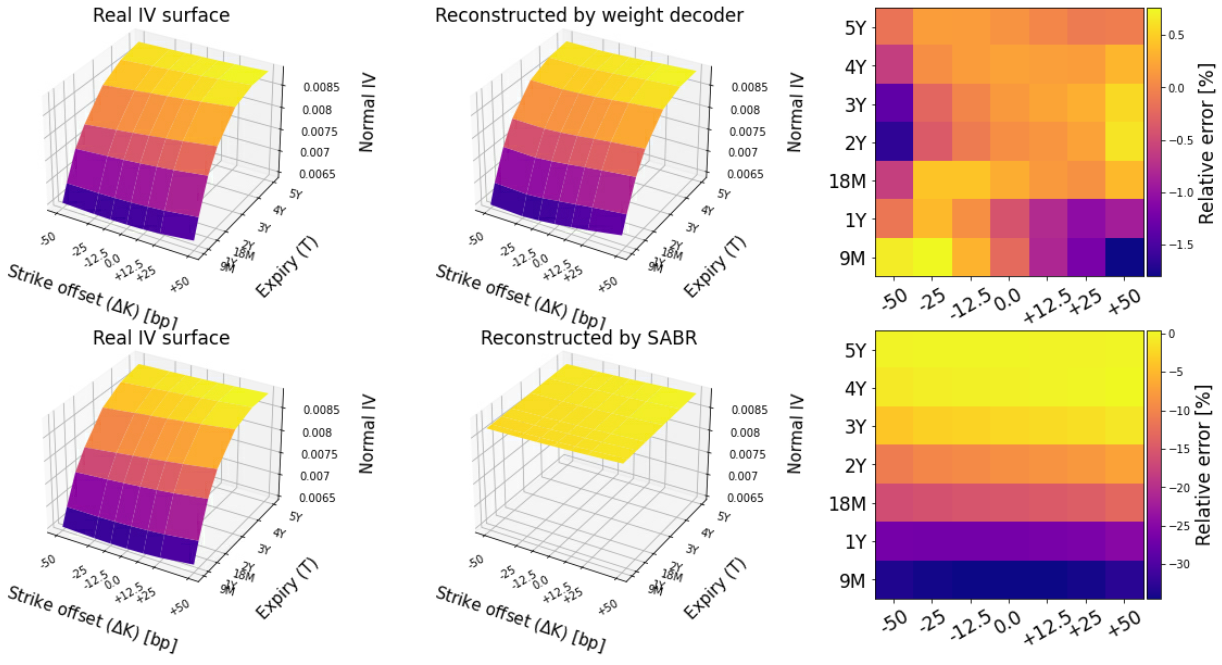}
\caption{Illustration of the implied volatility surface reproduction efficiency applying the weight decoder as well as the SABR model. From left to right we plotted the observed and reconstructed implied volatility surfaces as well as the relative error between them, respectively. Note, that the SABR model was calibrated on the last expiry.}
\label{fig:vols_sabr_probdec}
\end{figure}

In Figure \ref{fig:vols_sabr_probdec} we plotted the real and reconstructed volatility surfaces using the Weight Decoder neural network as well as the SABR model.

\begin{figure}[H]
\includegraphics[width=1.\linewidth]{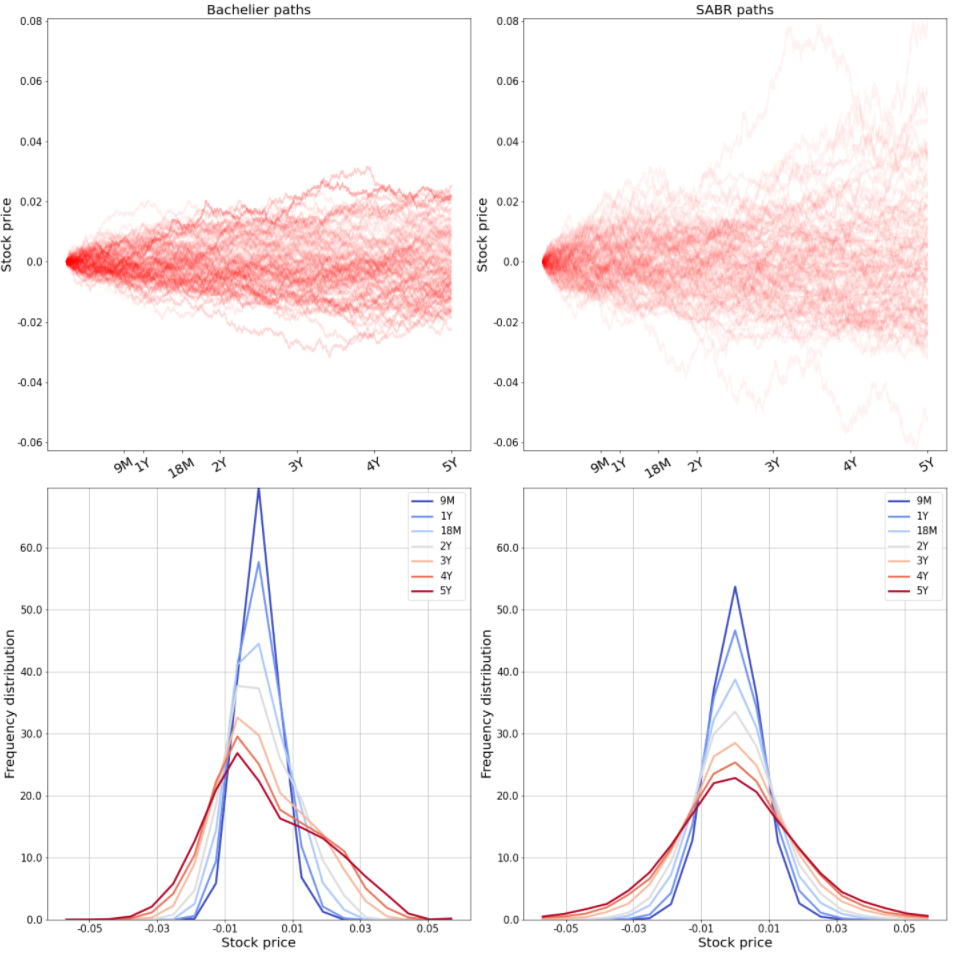}
\caption{\textit{Top row}: Stock price dynamics predicted by the Bachelier and the SABR models. In case of the Bachelier paths we marked the estimated weights of paths using the alpha value of the lines (more possible paths are less transparent). \textit{Bottom row}: Risk neutral densities at each expiry.}
\label{fig:bachelier_sabr_paths}
\end{figure}

We can see that the Weight Decoder is able to accurately reproduce the observed market with a mean relative absolute error of 0.41\%, while the SABR model fails to explain the volatilities of expiries shorter than 5 years, the discrepancy between the fitted and observed values can be more than 30\%. To illustrate the difference between the predicted dynamics of the underlying, we plot the related paths as well as the risk neutral distributions at each expiry in Figure \ref{fig:bachelier_sabr_paths}. We can notice that the weighted paths on the left hand side follow a slightly asymmetric probability density function, similar to a log-normal distribution. The paths related to the SABR model on the right hand side contain however more extreme realizations that can be seen in the wider probability densities as well.

\begin{figure}[H]
\includegraphics[width=1.\linewidth]{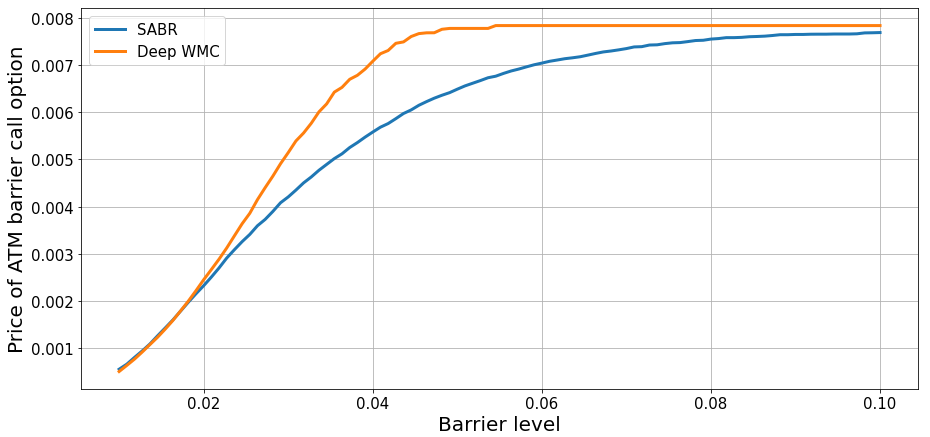}
\caption{Price of ATM barrier call option as a function of the barrier level. Prices have been calculated using i) weighted Bachelier and ii) unweighted SABR paths.}
\label{fig:barrier_option_price}
\end{figure}

Now, we will set the strike price of the barrier option to ATM and continuously modify the barrier from 0.01 to 0.1. The calculated option prices based on the two path sets can be seen in Figure \ref{fig:barrier_option_price}. We can see that as long as the barrier level is under 0.02 the prices are in agreement with each other. However, there is a significant difference in the prices in the range of [0.02, 0.08]. This implies that for such types of barrier options it is crucial to find the correct dynamics of the underlying asset. In our case the weighted Bachelier paths are able to calibrate to the term structure of the market, therefore the calculated barrier option prices are more reliable than using the single term SABR model.

\section{Discussion}\label{sec:discussion}

During our analysis we made two main findings: 1) We have confirmed that Variational Autoencoders are extremely effective in finding the correlations of the swaption market and hence they can massively reduce the number of degrees of freedom; 2) We demonstrated how to extract physically meaningful information from the derived latent space by translating the latent coordinates into weights of paths. We believe that our methodology can be used more generally. At many scientific fields we often face the problem of describing a specific phenomenon of the world by finding a mathematical model. Usually, as we observe more and more data the created analytic model is less and less powerful in explaining the new observations and therefore becomes more and more complex to keep its original performance. Contrarily, we claim that \textit{finding the minimally needed number of free parameters} for the first step of the analysis is very important and Variational Autoencoders are great tools in doing that. Creating an interpretable mathematical model with some model parameters makes sense only after the previous step. We can use then neural networks to find the non-trivial relations between the latent space and our model parameters. This final process essentially means to extract the physically meaningful information from the latent space.

\section{Conclusions}\label{sec:conclusions}

Our aim in this work was to find low dimensional representations of the swaption market. We built a Variational Autoencoder that was able to represent the market for more than five years using only three model parameters. We proposed a method to translate this static representation into a consistent dynamic model under the Weighted Monte Carlo framework initially staring from simple Brownian paths. Our Weight Decoder neural network was able to map the latent parameter space into the required path weights directly with good accuracy. The resulting (weighted) Monte Carlo measure can be applied to price non-vanilla derivatives such as barrier options. We compared prices to those resulting from a SABR model approach and demonstrated that the lack of matching the term structure of volatilities in the latter can lead to huge pricing differences. Clearly, finding a consistent dynamical model is crucial for adequate pricing of path dependent products. Our codes are publicly available at https://github.com/skunsagimate/DeepWMC.git.

\section{Acknowledgements}\label{acknowledgements}

Prepared with the support of the Doctoral Student Scholarship Program of the Cooperative Doctoral Program and the MILAB Artificial Intelligence National Laboratory Program of the Ministry of Innovation and Technology from the National Research, Development and Innovation Fund.

\section{Declaration of interest}

The authors report no conflicts of interest. The authors alone are responsible for the content and writing of the paper.

%\newpage

\bibliographystyle{apalike} 
\bibliography{references}

\end{document}